\documentclass{nature_compact}

\usepackage{epsfig}
\usepackage{nature_macros}
\usepackage{amssymb}

\title{The characteristic blue spectra of accretion disks in quasars\\
  as uncovered in the infrared}

\author{Makoto Kishimoto$^{1,2}$, Robert Antonucci$^{3}$, Omer
Blaes$^{3}$, Andy Lawrence$^{2}$, Catherine Boisson$^{4}$, Marcus
Albrecht$^{5}$ \& Christian Leipski$^{3}$\\}

\begin{document}

\maketitle

\begin{affiliations}
 \item Max-Planck-Institut f\"ur Radioastronomie, Auf dem H\"ugel 69,
 53121 Bonn, Germany
 \item SUPA (Scottish Universities Physics Alliance), Institute for
 Astronomy, University of Edinburgh, Royal Observatory, Blackford
 Hill, Edinburgh, EH9 3HJ, UK
 \item Physics Department, University of California, Santa Barbara, CA
 93106, USA
 \item LUTH, FRE 2462 du CNRS, associ\'ee \`a l'Universit\'e Denis Diderot,
    Observatoire de Paris, Section de Meudon, 
    F--92195 Meudon Cedex, France
 \item Instituto de Astronom\'ia, Universidad Cat\'olica del Norte
   (UCN), Avenida Angamos 0610, Antofagasta, Chile
\end{affiliations}

\renewcommand{\thefootnote}{\fnsymbol{footnote}}

\begin{abstract}

Quasars are thought to be powered by supermassive black holes
accreting surrounding gas\cite{Salpeter64,Lynden-Bell69,Shields78}.
Central to this picture is a putative accretion disk which is believed
to be the source of the majority of the radiative
output\cite{Lynden-Bell69,Shields78,Malkan83may}.  It is well known,
however, that the most extensively studied disk model\cite{Shakura73}
--- an optically thick disk which is heated locally by the dissipation
of gravitational binding energy --- is apparently contradicted by
observations in a few major respects\cite{Antonucci99,Koratkar99}. In
particular, the model predicts a specific blue spectral shape
asymptotically from the visible to the
near-infrared\cite{Shakura73,Hubeny00}, but this is not generally seen
in the visible wavelength region where the disk spectrum is
observable\cite{Neugebauer87,Cristiani90,Francis91,Zheng97,VandenBerk01}.
A crucial difficulty was that, toward the infrared, the
disk spectrum starts to be hidden under strong hot dust emission from
much larger but hitherto unresolved scales, and thus has essentially
been impossible to observe.  Here we report observations of polarized
light interior to the dust-emiting region that enable us to uncover
this near-infrared disk spectrum in several quasars.  The revealed
spectra show that the near-infrared disk spectrum is indeed as blue as
predicted.  This indicates that, at least for the outer
near-infrared-emitting radii, the standard picture of the locally
heated disk is approximately correct.


\end{abstract}

A success of the most extensively studied disk model is that it gives
the radiative output peak approximately correctly in the ultraviolet
(UV; $\sim$0.01-0.4 $\mu$m) wavelengths for the case of a supermassive
black hole. This is observed for a generic spectral energy
distribution\cite{Sanders89} of quasars, the most luminous example of
active galactic nuclei (AGN).  However, it has long been known that
the model apparently shows a few major contradictions with
observations\cite{Antonucci99,Koratkar99}.  One of the disagreements,
and perhaps the most easily comprehensible one, is the spectral shape
of the radiation.  From the basic hypothesis of the model (an optically
thick disk heated locally), the effective disk temperature $T$ is
fixed as a function of radius $r$ as $T \propto r^{-3/4}$ over a broad
range of radii.  This leads to a well-known blue spectral shape limit,
$F_{\nu} \propto \nu^{+1/3}$, being asymptotically reached at long
wavelengths from the visible (also called the optical; $\sim$0.4-1
$\mu$m) to the near-infrared ($\sim$1-2 $\mu$m) for AGN disks.  In
contrast, many studies have shown that the general AGN spectral shape
observed at optical/UV wavelengths is much redder, spectral slope
$\alpha$ being from $-$0.2 to $-$1 ($F_{\nu} \propto \nu^{\alpha})$,
and never as blue as this spectral
shape\cite{Neugebauer87,Cristiani90,Francis91,Zheng97,VandenBerk01}.

The predicted blue shape limit is strictly true in the simplest
assumption of local black-body emission. In more sophisticated disk
atmosphere models\cite{Hubeny00}, the spectrum generally becomes
slightly redder at optical wavelengths, owing to various opacity and
non-LTE effects, but discrepancies between the model and observed
spectra still remain\cite{Davis07}. However, the redder model slopes
at optical wavelengths form a wider concave spectrum that shifts the
bluer limit above to longer wavelengths, into the near-IR.  The
observed spectra certainly appear to become bluer from the short UV
wavelengths to the optical. The crucial observational difficulty here
has been that the disk spectrum starts to be hidden under the hot dust
thermal emission which begins at wavelengths greater than $\sim$1
$\mu$m, a limit set by the sublimation temperature of dust grains
($\sim$1500 K).  These dust grains exist at larger spatial scales, in
a configuration often thought to have a torus-like geometry but which
is generally not yet spatially resolvable. Therefore, it has been
virtually impossible to observe the underlying near-IR disk
spectrum\cite{Malkan89}. We note that, contrary to early spectral
fitting studies\cite{Shields78,Malkan83may,Malkan83dec,Malkan89}, the
IR component is no longer thought to be non-thermal, and thus cannot
be extrapolated to underlie the optical spectrum --- an extrapolated
non-thermal spectrum had effectively made the inferred disk spectrum
on top of the non-thermal spectrum bluer.

We argue here that this buried part of the disk spectrum can be
revealed by observing the near-IR polarized light.  Optical continua
of many directly-visible AGNs called Type 1s (Seyfert 1 galaxies
and quasars) are known to be linearly polarized at a level
$\lesssim$1\%.  The polarization position angle (PA) in these Type 1
cases is mostly parallel to the rotation axis of the putative
accretion disk, where the axis can be probed by the linear jet-like
structure of radio
emission\cite{Antonucci83may,Smith02,Smith04may}. (This is in contrast
to the cases in hidden AGNs or Type 2s, which show high polarization
at perpendicular PAs and strong broad lines in polarized light, and
which are not the subject of the present paper.)  This polarization in
Type 1s is interpreted as an indication of an equatorial scattering
region, which is optically thin and surrounds the disk.  In many
Seyfert 1 galaxies, broad emission lines are polarized at a much lower
level than the continuum polarization and at different
PAs\cite{Smith02,Smith04may}, indicating that the scatterers reside
roughly at the same spatial scales as the broad-line-emitting
clouds. 

At least in several quasars, the emission line polarization even
vanishes --- the optical polarized light spectrum shows no or very
little emission line flux\cite{Kishimoto03,Kishimoto04}. This is very
likely to indicate that the scatterers reside interior to the
broad-line clouds.  In these Type 1 cases, the scatterers are thought
to be electrons and not dust grains, since the scattering region is
interior to the dust sublimation radius.  Because electron scattering
is wavelength independent, the polarized light therefore produces a
copy of the spectrum originating interior to the scattering region.
(We note that this electron scattering is conceptually different from
the one in previous works\cite{Webb93,Impey95}, which was assumed to
be intrinsic to the accretion disk atmosphere and gave rise to the
prediction of perpendicular PAs.)  In the optical polarized light from
these quasars, which excludes the emission from the broad-line region,
we actually found a hydrogen Balmer-edge feature in absorption for the
first time, which we believe originates in the disk and is a very
specific indication of the thermal and optically-thick nature of the
emission\cite{Kishimoto03,Kishimoto04}.

\begin{figure}[th]
\epsfig{file=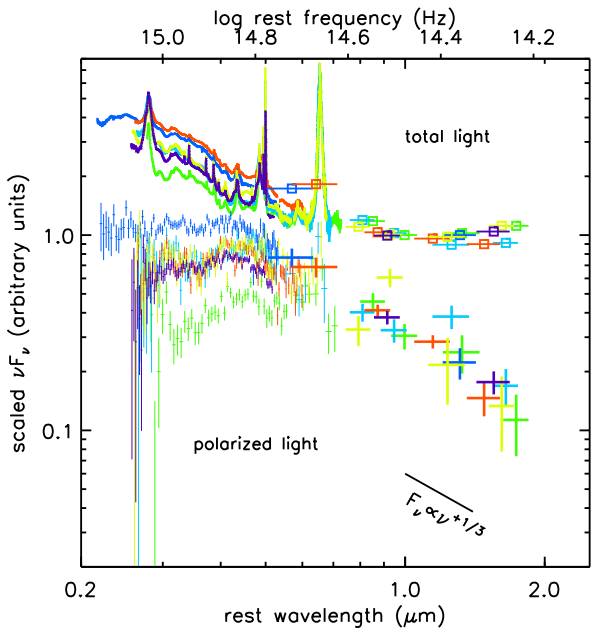,width=89mm}

\caption{{\small {\bf Overlay of the polarized- and total-light
  spectra observed in six different quasars.} We plot scaled $\nu
  F_{\nu}$ data: Q0144-3938 (redshift $z$=0.244), green; 3C95
  ($z=$0.616), blue; CTS A09.36 ($z$=0.310), light blue; 4C 09.72
  ($z$=0.433), red; PKS 2310-322 ($z$=0.337), light green. Plotted in
  purple are the data for Ton 202 ($z$=0.366) from a previous
  paper\cite{Kishimoto05}.  Total-light spectra, shown as bold traces
  in the optical and as squares in the near-infrared, are normalized
  at 1$\mu$m in the rest frame, by interpolation (except for 3C95 for
  which we normalized by $\nu F_{\nu}$ observed at 1.3 $\mu$m in the
  rest frame).  Polarized-light spectra, shown as light points in the
  optical and as bold points in the near-infrared both (vertical error
  bars, 1-$\sigma$), are separately normalized, also at 1$\mu$m, by
  fitting a power-law to the near-IR polarized-light spectra.  For
  both total-light and polarized-light data, horizontal bar lengths
  indcate bandwidth.  The normalized polarized-light spectra are
  arbitrarily shifted downwards by a factor of three relative to the
  normalized total-light spectra, for clarity. The total-light spectra
  in $\nu F_{\nu}$ turns up at around or slightly longward of 1
  $\mu$m.  In contrast, the polarized-light spectra in $\nu F_{\nu}$
  all consistently and systematically decrease towards long
  wavelengths, showing a blue shape of approximately power-law
  form. }}

\label{fig-overlay}
\end{figure}

We can then use the same polarized light, but in the near-IR, to
reveal the hidden spectrum of the disk, by stripping off the dust
radiation from the torus exterior to the scattering region.  In
previous work\cite{Kishimoto05}, we suggested that this method appears
to work in at least one quasar that has polarization data with high
signal-to-noise-ratio at two wavelength bands in the near-IR.
However, crucial information needed at the time was whether the
near-IR polarized light behavior is consistent and systematic in
different objects. If it is, this would critically argue against, for
example, a possible secondary polarization component related to dust
grains newly arising in the near-IR.  Therefore we undertook the
near-IR polarimetry of five other quasars. The targets were selected
to be polarized in optical continua but essentially {\it not} in
emission lines, to ensure the scattering to be interior to the
broad-line region.  These properties were either already
known\cite{Kishimoto04} or were determined in our optical polarimetric
survey and follow-up spectropolarimetry.

In Fig.1 we show the spectra of linearly polarized light measured in
the near-IR broad-band imaging polarimetry, and optical
spectropolarimetry, of the six quasars in total (including the one
studied in the previous work).  Details of the measurements, as well
as the procedures for removing instrumental polarizations, can be
found in the Supplementary Information. Generally, polarization
degrees observed for these quasars are $\sim$1\% at $\sim$0.5 $\mu$m,
gradually decreasing to $\sim$0.5\% at $\sim$2 $\mu$m, and PAs are
essentially constant over the near-UV/optical/near-IR wavelengths for
a given object.  A significant result here is that all the objects
behave in a similar and systematic way showing blue polarized-light
spectra.  While the total-light spectra in $\nu F_{\nu}$ turn up at
around 1 $\mu$m toward longer wavelengths due to the onset of dust
emission, all the polarized light spectra, which eliminate dust,
display a rapid decrease in $\nu F_{\nu}$, with a shape of
approximately power-law form.

The measurement of the spectral index $\alpha$ in $F_{\nu}$ ($\propto
\nu^{\alpha}$) for each object is shown in Fig.2.  The measured slopes
are consistent with each other within their errors, and the individual
slopes as well as their average clearly point to a shape much bluer
than those observed in the UV/optical. Astonishingly, they are all
consistent with the $F_{\nu} \propto \nu^{+1/3}$ shape.  The weighted
mean of the measured slopes is $\alpha = +0.44\pm0.11$.  Although the
sample size is small, there does not appear to be any luminosity
dependence, as seen in Fig.2, and we did not find dependencies on
black hole masses $M_{\rm BH}$ or Eddington ratios $L/L_{\rm Edd}$
derived from the width of the H$\beta$ Balmer line.  This is expected
if the near-IR spectrum is in the long-wavelength limit of the disk
model, which is independent of parameters such as black hole mass or
Eddington ratio.  In this case, by regarding each measurement as a
measurement of the same quantity, the weighted mean over these
measurements becomes physically meaningful.  We note that, if we
formally convert the mean slope to the radial temperature distribution
for the case of an optically-thick disk, we obtain $T \propto
r^{-0.78\pm0.03}$, consistent with the predicted dependence $T \propto
r^{-3/4}$.

\begin{figure}
\epsfig{file=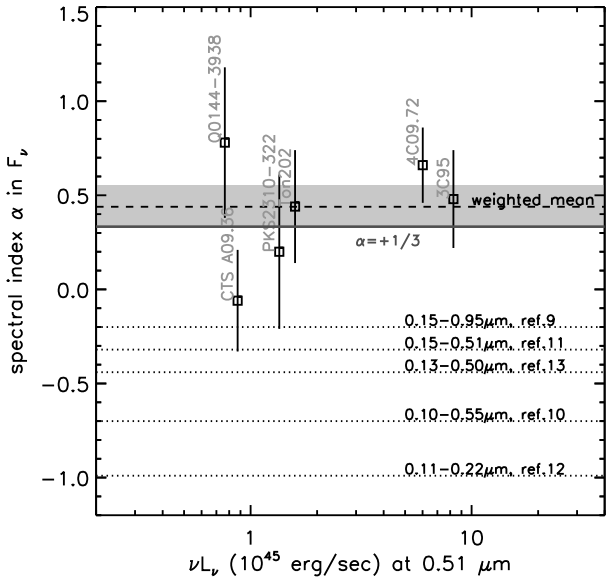,width=89mm}
\caption{{\small {\bf Spectral index of polarized light spectra}. We plot
  $\alpha$ (in $F_{\nu} \propto \nu^{\alpha}$) against $\nu L_{\nu}$
  for total light at 0.51 $\mu$m. The index was measured by a
  power-law fit for each near-IR polarized-light spectrum (note the
  different wavelength range covered depending on the redshift) and is
  shown with 1-$\sigma$ error bars. A weighted mean of the spectral
  index measurements is shown dashed; the shaded area represents its
  deduced 1-$\sigma$ uncertainty.  The mean or median slopes of the
  UV/optical total-light spectra derived in various
  studies\cite{Neugebauer87,Francis91,VandenBerk01,Cristiani90,Zheng97}
  are also shown.}}
\label{fig-slope}
\end{figure}

The systematic behavior of the near-IR polarized light, as well as the
constancy of the PAs over all wavelengths, strongly argues against
there being any secondary polarization contamination.  We might worry
that the polarized light would be affected if the corresponding
spatial scale of the disk emission at long wavelengths were to become
large and finite in comparison with the size of the scattering region.
However, this seems unlikely, since the half-light radius of the disk
(within which half of the total light is emitted) even at 2 $\mu$m is
still much smaller ($\sim$400$R_S$, where $R_S=2 G M_{\rm BH}/c^2$)
than at least the radius of the broad-line region\cite{Bentz06}
($\sim$4000$R_S$), in the case of an untruncated multi-temperature
black-body disk for our quasars.  Significant geometrical effects will
not occur unless the disk emission size becomes almost the same as the
scattering region size.  Therefore the near-IR polarized light spectra
very likely reveal the intrinsic accretion disk spectra.

The measured slopes, being as blue as the slope of the predicted shape
$F_{\nu} \propto \nu^{+1/3}$, strongly suggest that, at least in the
outer near-IR emitting radii, the standard but hitherto unverified
picture of the disk being optically thick and locally heated is
approximately correct. In this case, an implication is that
other model problems at shorter wavelengths are associated with,
or originate from, our lack of understanding of the inner regions of
the same disks.  We note that disk irradiation in limiting cases can
contribute to the heating without changing the $\nu^{+1/3}$ spectral
shape at long wavelengths; but it would not dominate the local
internal heating in the outer radii considered here, and thus is not
directly relevant except for some very specific
cases\cite{Blaes04,Agol00}.

The standard disk is also well known to be gravitationally unstable at
large radii\cite{Shlosman87}. These radii may well correspond to those
emitting in the IR\cite{Goodman03} ($\sim$800$R_S$ for our
quasars). In this case, if the disk is truncated at such a radius, the
spectrum will show a break, becoming even bluer at the longest
wavelengths\cite{Kishimoto05}.  Although statistically insignificant,
our data do suggest that the near-IR slope is slightly bluer than the
spectral shape $F_{\nu} \propto \nu^{+1/3}$, with a hint of possibly
becoming bluer at longer wavelengths.  This can be followed up by
extending the wavelength coverage with similar polarized Type-1 AGNs
at lower redshifts.  Such future measurements may pioneer the way to
probe how and where the disk ends and how material is being supplied
to the nucleus.

\bigskip





\begin{addendum}

 \item[Supplementary Information] is linked to the online version of
 the paper at www.nature.com/nature.

 \item The United Kingdom Infrared Telescope is operated by the Joint
 Astronomy Centre on behalf of the Science and Technology Facilities
 Council of the U.K.  We thank the Department of Physical Sciences,
 University of Hertfordshire, for providing IRPOL2 for the UKIRT. This
 research is partially based on observations collected at the European
 Southern Observatory, Chile. 

 \item[Author Information] Correspondence and requests for materials
should be addressed to M.K.~(email: mk@mpifr-bonn.mpg.de).
\end{addendum}


\end{document}